\documentclass[aps,twocolumn,a4paper,showpacs]{revtex4}
\usepackage{graphicx}
\usepackage{color}
\usepackage{amsmath}
\usepackage{amssymb}
\usepackage{enumerate}
\newcommand{\be}{\begin{equation}}
\newcommand{\ee}{\end{equation}}
\newcommand{\ben}{\begin{eqnarray}}
\newcommand{\een}{\end{eqnarray}}
\newcommand{\bes}{\begin{subequations}}
\newcommand{\ees}{\end{subequations}}

\newcommand{\bb}{\bibitem}

\newcommand{\vphi}{\varphi}
\begin{document}
\title{Compact Q-balls}
\author{D. Bazeia$^{1}$, L. Losano$^{1}$,  M.A. Marques$^1$, R. Menezes$^{2,3}$, and R. da Rocha$^4$}
\affiliation{$^1$Departamento de F\'\i sica, Universidade Federal da Para\'\i ba, 58051-970 Jo\~ao Pessoa, PB, Brazil}
\affiliation{$^2$Departamento de Ci\^encias Exatas, Universidade Federal da Para\'{\i}ba, 58297-000 Rio Tinto, PB, Brazil}
\affiliation{$^3$Departamento de F\'\i sica, Universidade Federal de Campina Grande, 58109-970 Campina Grande, PB, Brazil}
\affiliation{$^4$Centro de Matem\'atica, Computa\c c\~ao e Cogni\c c\~ao, Universidade Federal do ABC, 09210-580 Santo Andr\'e,  Brazil}
\begin{abstract}
In this work we deal with non-topological solutions of the Q-ball type in two space-time dimensions, in models described by a single complex scalar field that engenders global symmetry. The main novelty is the presence of stable Q-balls solutions that live in a compact interval of the real line and appear from a family of models controlled by two distinct parameters. We find analytical solutions and study their charge and energy, and show  how to control the parameters to make the Q-balls classically and quantum mechanically stable.
\end{abstract}
\date{\today}
\maketitle
\section{Introduction}

This work deals with Q-balls, which are defect structures of the non-topological type. As one knows, defect structures can be of topological or non-topological nature. The topological structures are stable because one can associate conserved topological currents to them, and they are conserved due to the topological properties of the solutions \cite{b1}. However, the non-topological structures \cite{b2,tdlee} do not have associated topological currents, so they are unstable, although they can be stabilized in a diversity of cases, in particular as Q-balls \cite{coleman,kolb}. 

The case of Q-balls is of current interest, and in the simplest situation one can deal with a single complex scalar field. 
Some key properties of Q-balls have been largely studied in the literature \cite{tdlee,coleman,kolb,41,41a,prl,42,kusenko,dm,q1,tuomas,minos,q2,sut,ku,r1,r2,sut1,ed,sta1,sta2,bmm}. In general, Q-balls appear related to the existence of global U(1) symmetries, in the Standard Model the presence of global U(1) symmetries can be related to fermionic (baryonic, leptonic) charges, and in extended supersymmetric models the scalar superpartners of baryons and leptons can condensate and give rise to Q-balls. In this sense, Q-balls are of current interest to both baryogenesis and leptogenesis. Possibilities of formation of Q-balls, during a phase transition in the early universe, have been studied as a part of solitogenesis \cite{kolb}, and can contribute to dark matter, in the present era of the universe \cite{kus1}. Q-balls experimental detection has been proposed in this context, having  a parameter region different
from that for gravitino dark matter in  IceCube \cite{ice}.
 As it is known, one can suggest that the particle asymmetry of the universe appears as in the Affleck-Dine mechanism \cite{ad}, as a feature of the flat direction condensate which gives rise to Q-balls \cite{q1,q2}.

In a recent work \cite{bmm}, some of us studied Q-balls in models that allow the presence of analytic solutions, investigating features such as shape, energy, charge, stability and splitting of the solutions. In the current work we go further on the subject and we modify the potential, to obtain a new kind of Q-balls, of the compact type.
As one knows, defect structures having compact support were first studied in \cite{RH}, in the context of integrable models, but the  motivation for the current work comes from the recent works on defect structures of the compact type \cite{c,cl}. In particular, in Ref.~\cite{cl} the presence of compact lumps was investigated, and this motivated us to go further to construct compact Q-balls. We recall that compact Q-balls appeared before in \cite{aro}, with the scalar potential being of the
signum-Gordon type, having a V-shaped form. The approach here is different, and we introduce a new family of models, with the potential controlled by two distinct parameters, as we explain in Sec.~III. 

The investigation starts in Sec.~\ref{sec:gen}, where we describe a simple model and review some basic facts about Q-balls. We continue the study in Sec.~\ref{sec:c}, where we investigate a new family of models, described by a potential which is controlled by two distinct parameters. There, we show that the new models support analytical solutions of the compact type, which we use to describe the charge and energy, and to study stability. Since the compact structure does not have a tail, it  seems to behave as hard charged ball, well different from the standard Q-balls. We summarize the results and add some comments to end the work in Sec.~\ref{sec:end}.

\section{Generalities} 
\label{sec:gen}

In order to investigate Q-balls, we consider the Lagrange density
\be\label{lgeneral}
{\cal L} = \frac12 \partial_\mu{\bar\vphi} \partial^\mu \vphi - V(|\vphi|),
\ee
where $\vphi$ is a complex scalar field and $V(|\vphi|)$ is the potential. We are working in $(1,D)$ spacetime dimensions, so the equation of motion for $\varphi$ has the form
\be
\ddot{\vphi} - \nabla^2 \vphi + \frac{\vphi}{|\vphi|} \frac{dV}{d|\vphi|} = 0.
\ee
To search for Q-balls we take the usual {\emph{ansatz}}
\be\label{ansatz}
\vphi(r,t)=\sigma(r)\,e^{i\omega t}.
\ee
The conserved Noether charge is
\be\label{charge}
Q = \frac{1}{2i} \int_{-\infty}^\infty{d^Dx\left({\bar\vphi}\dot{\vphi} - \vphi\dot{\bar\vphi}\right)}=
\omega \int_{-\infty}^\infty{d^Dx\, \sigma^2(r)}.
\ee
The equation of motion becomes 
\be\label{eom}
\sigma^{\prime\prime} = \frac{1-D}{r}\sigma^{\prime} -\omega^2\sigma + \frac{dV}{d\sigma}.
\ee
As usual, we consider the boundary conditions
\ben\label{bcond}
\sigma^\prime(0) = 0;\;\;\;\;\;\sigma(\infty) = 0.
\een
The above equation of motion \eqref{eom} can be seen in the form
\be\label{eqeff}
\sigma^{\prime\prime} = \frac{1-D}{r}\sigma^{\prime} + \frac{dU}{d\sigma},
\ee 
with $U=U(\sigma)$ being a kind of effective potential for the field $\sigma$. It has the form
\be\label{veff}
U(\sigma) = V - \frac12\omega^2\sigma^2.
\ee

As one knows, in order to have solutions obeying the boundary conditions \eqref{bcond}, $\omega$ must be in the interval
\be\label{condomega1}
\omega_-<\omega<\omega_+,
\ee
with $\omega_+ = V^{\prime\prime}(0)$ and $\omega_-=\sqrt{2V(\sigma_0)/\sigma_0^2}$, where $\sigma_0$ is the minimum of $V(\sigma)/\sigma^2$. We call $\omega_+$ the upper bound and $\omega_-$ the lower bound for the frequency. 
The energy-momentum tensor for this model is
\be\label{emt}
T_{\mu\nu} = \frac12\partial_\mu{\bar\vphi}\partial_\nu\vphi + \frac12\partial_\mu\vphi\partial_\nu{\bar\vphi} - \eta_{\mu\nu}{\mathcal L}.
\ee
The energy density can be calculated from the $T_{00}$ component of Eq.~\eqref{emt} with the Lagrange density given by
Eq.~\eqref{lgeneral}. It is given by, changing $T_{00}\;{\mapsto}\; \epsilon$,
\be
\epsilon= \epsilon_k + \epsilon_g + \epsilon_p,
\ee
which represents the kinetic, gradient and potential energy densities, respectively. They are given by
\ben
\epsilon_k  &=& \frac12|\dot{\vphi}|^2, \\
\epsilon_g &=& \frac12|\vphi^{\prime}|^2, \\
\epsilon_p &=&  V(|\vphi|)
\een
By using the {\emph{ansatz}} \eqref{ansatz}, the energy density becomes
\be\label{energydens1}
\epsilon = \frac{1}{2}\omega^2\sigma^2 + \frac12{\sigma^\prime}^2 + V(\sigma).
\ee
When integrated, it gives the total energy of the Q-ball. We can see that the kinetic energy can be written in terms of the conserved charge as
\be\label{ke}
E_k=\frac12 \omega Q.
\ee
The other components of the energy-momentum tensor \eqref{emt} are
\ben
T_{01} &=& Re\left(\dot{\bar\vphi} \vphi^\prime\right), \\
T_{11} &=& \frac12|\vphi^\prime|^2 + \frac12|\dot{\vphi}|^2 -V(|\vphi|).\;\;\;
\een
With the {\emph{ansatz}}
 \eqref{ansatz}, they become
\ben
T_{01} &=& 0, \\
T_{11} &=& \frac12{\sigma^\prime}^2 - V(\sigma).
\een
Since the energy-momentum tensor is conserved, i.e., $\partial_\mu T^{\mu\nu}=0$, we see that $T_{11}$ is constant. In particular, if the solution satisfies the condition $T_{11}=0$, the energy densities (as well as the energies), are related by
\be
\epsilon_p = \epsilon_k + \epsilon_g,
\ee
and we only have to know the kinetic and gradient portions to find the total energy density. This allows us to write the total energy as
\be
E = 2(E_k+E_p).
\ee

We now turn our attention to the stability of the Q-balls and consider the following two types of stabilities \cite{sta2}: 
\begin{enumerate}[(a)]
\item The first case is the quantum mechanical stability, that is the stability with respect to decay into free particles. As it was stated in Eq.~\eqref{condomega1}, Q-ball solutions exist for $\omega$ in a specific range of values. The Q-ball is stable if the ratio between the energy and the charge satisfies $E/Q < \omega_+$. 

\item The second type of stability is known as the classical stability, which is the one that concerns small perturbations of the field. The Q-ball is classically stable if $dQ/d\omega<0$. This means that the charge $Q$ is monotonically decreasing with $\omega$.
As non-trivial configurations in scalar field theory having unbroken global symmetry, Q-balls, satisfying this constraint, are thus solutions to the equation of motion for Noether charge $Q$, being hence spatially localized, stable, solitons.
\end{enumerate}
There is another type of stability, against fission, which requires $d^2E/dQ^2<0$. However, as we know that $\partial E/\partial Q=\omega$, it is straightforward to show that classically stable Q-balls are also stable against fission.

It is worth to mention that, irrespectively of the type of stability, Q-balls are hence stable soliton solutions, rotating in an internal symmetry space, differently of static soliton solutions, as vortices,   monopoles, kinks and  skyrmions, among other solutions \cite{Otsu:1988bn}. 
Let us now study Q-balls in $(1,1)$ spacetime dimensions, searching for analytical solutions. We consider the model
\be\label{lreview}
{\cal L} = \frac12 \partial_\mu\vphi^* \partial^\mu \vphi - \frac12 |\vphi|^2 + \frac13|\vphi|^3 - \frac14 a\,|\vphi|^4.
\ee
We take the usual {\emph{ansatz}}, and this leads us to the effective potential
\be\label{veff1}
U(\sigma) = \frac12 (1-\omega^2)\sigma^2 -\frac13 \sigma^3 + \frac14 a\,\sigma^4.
\ee
This model was studied in \cite{41} and the solution can be written as \cite{41,bmm}
\ben\label{sol1}
\sigma(x) &=& \sqrt{\frac{1-\omega^2}{2a}}\left[\tanh\left(\frac12 \sqrt{1-\omega^2}\, x + b \right) \nonumber\right. \\
&&\left.-\tanh\left(\frac12 \sqrt{1-\omega^2}\, x - b \right)\right],
\een
where
\be
b=\frac12 \text{arctanh}\left({3\sqrt{(1-\omega^2)a/2}}\right).
\ee
The expression in Eq.~\eqref{sol1} is an exact solution of the equation of motion \eqref{eom}, that obeys the boundary conditions \eqref{bcond} for several values of $\omega$. For $a\geq2/9$, $\omega$ is bounded according to \eqref{condomega1}, with $\omega_-={1-2/(9a)}$ and $\omega_+=1$. We note that $b$ controls the shape of the solution, which is bell-shaped.

\begin{figure}[t!]
\includegraphics[width=6 cm,height=5.5cm]{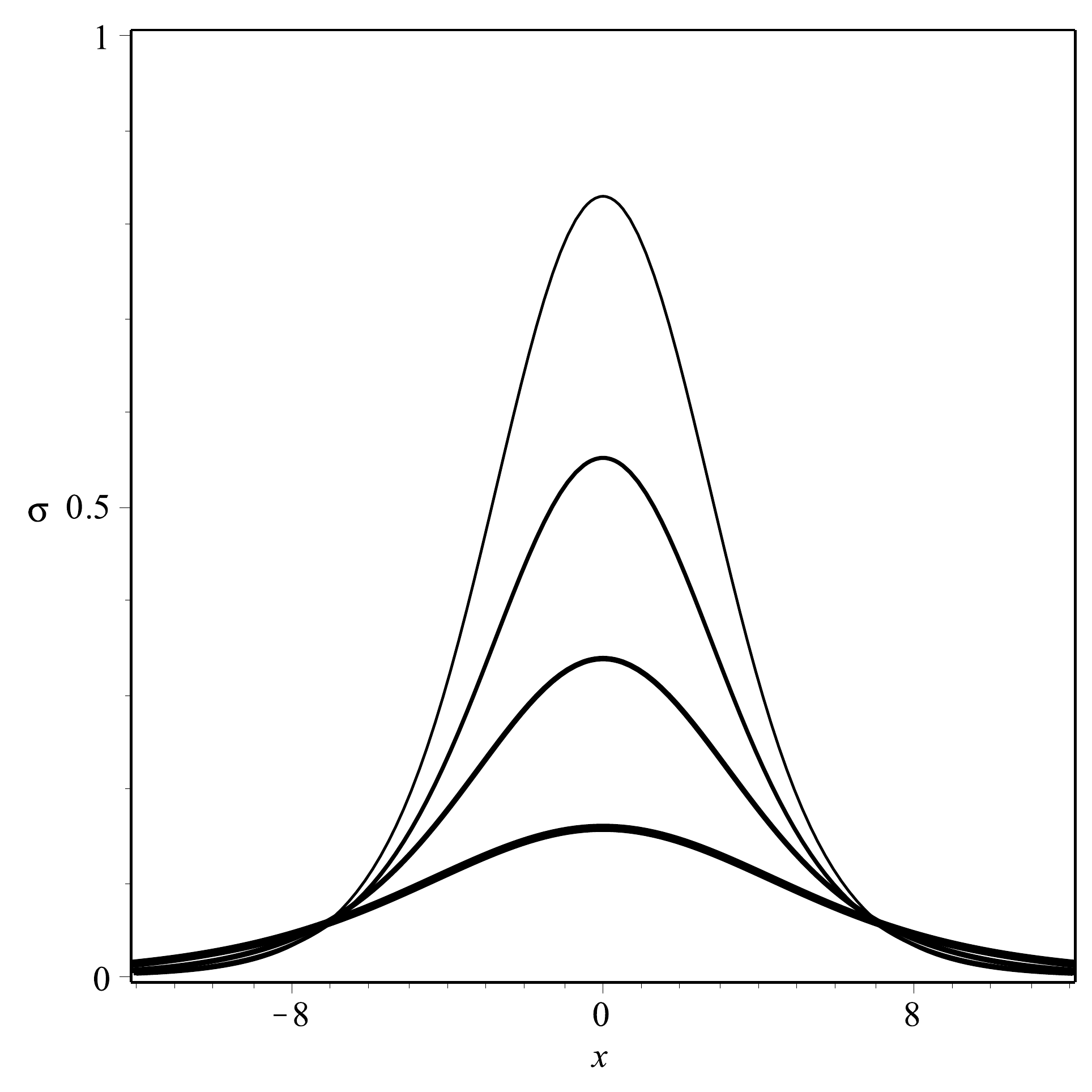}
\caption{The solution given by \eqref{sol1}, depicted for $a=4/9$ and $\omega=0.6,0.7,0.8$, and $0.9$. The thickness of the lines increases as $\omega$ increases.}\label{fig1}
\end{figure}

It is possible to use the exact solution \eqref{sol1} to calculate the charge from Eq.~\eqref{charge}. We get
\be\label{charge1}
Q=\frac{4\omega\sqrt{1-\omega^2}}{a} \left(2b\coth(2b)-1\right).
\ee
In the recent investigation \cite{bmm} it was shown that the solution \eqref{sol1} is stable for $a>0.2253973$, and in Fig.~\ref{fig1} we depict the solution \eqref{sol1} for some values of $a$ and $\omega$. The profile of the solutions shows the standard situation, with the field vanishing asymptotically, exponentially. 

The behavior shown in Fig.~\ref{fig1} should be contrasted with the behavior of the new compact Q-balls which we study in the next Section. The study is motivated by the recent work \cite{cl}, which shows that it is possible to construct non-topological solutions that lives in a compact space. With this on mind, in the next Section we investigate models that admit the existence of compact Q-balls.

\section{Compact Q-Balls}
\label{sec:c}
\begin{figure}[t!]
\includegraphics[width=6cm,height=5.5cm]{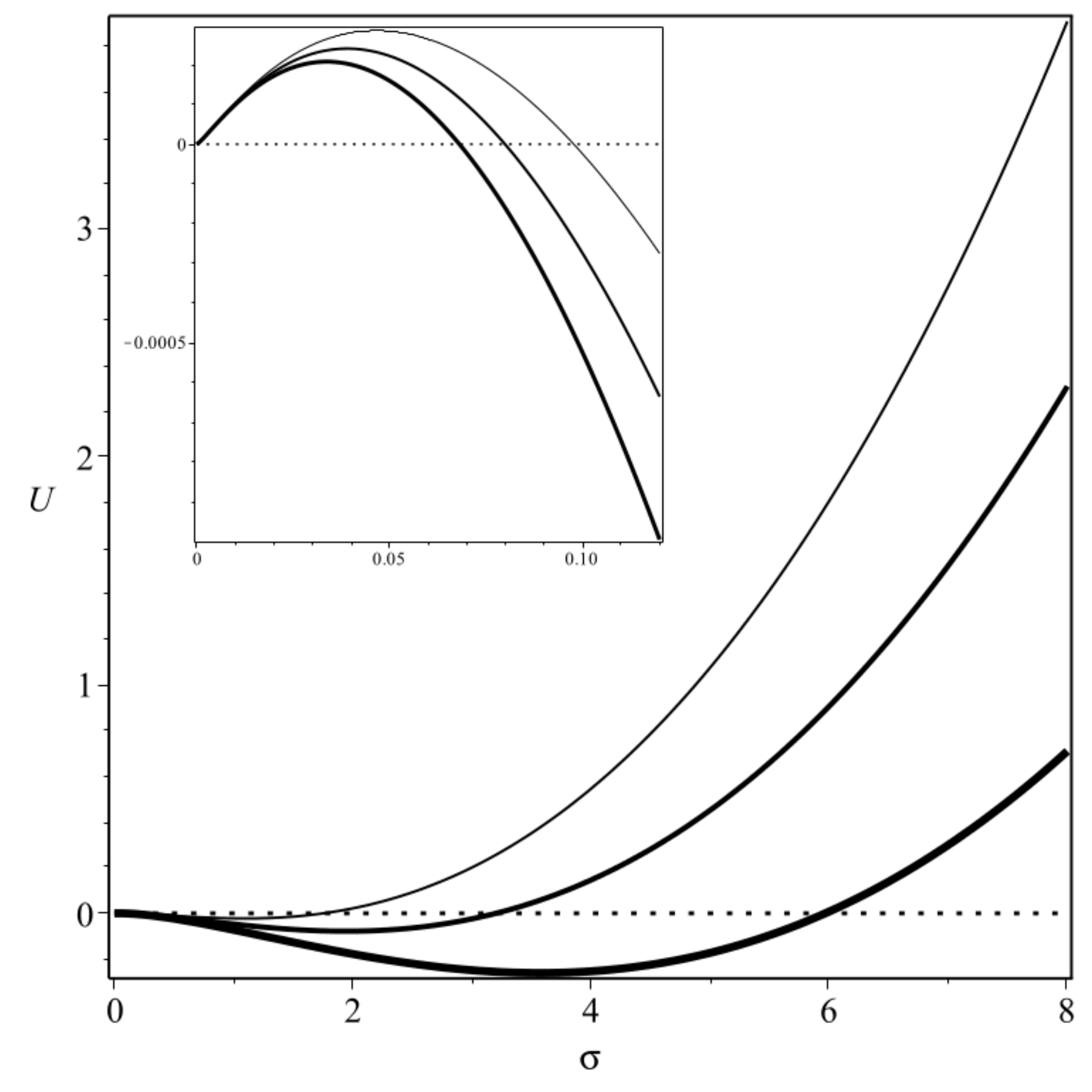}
\includegraphics[width=6.4cm,height=5.5cm]{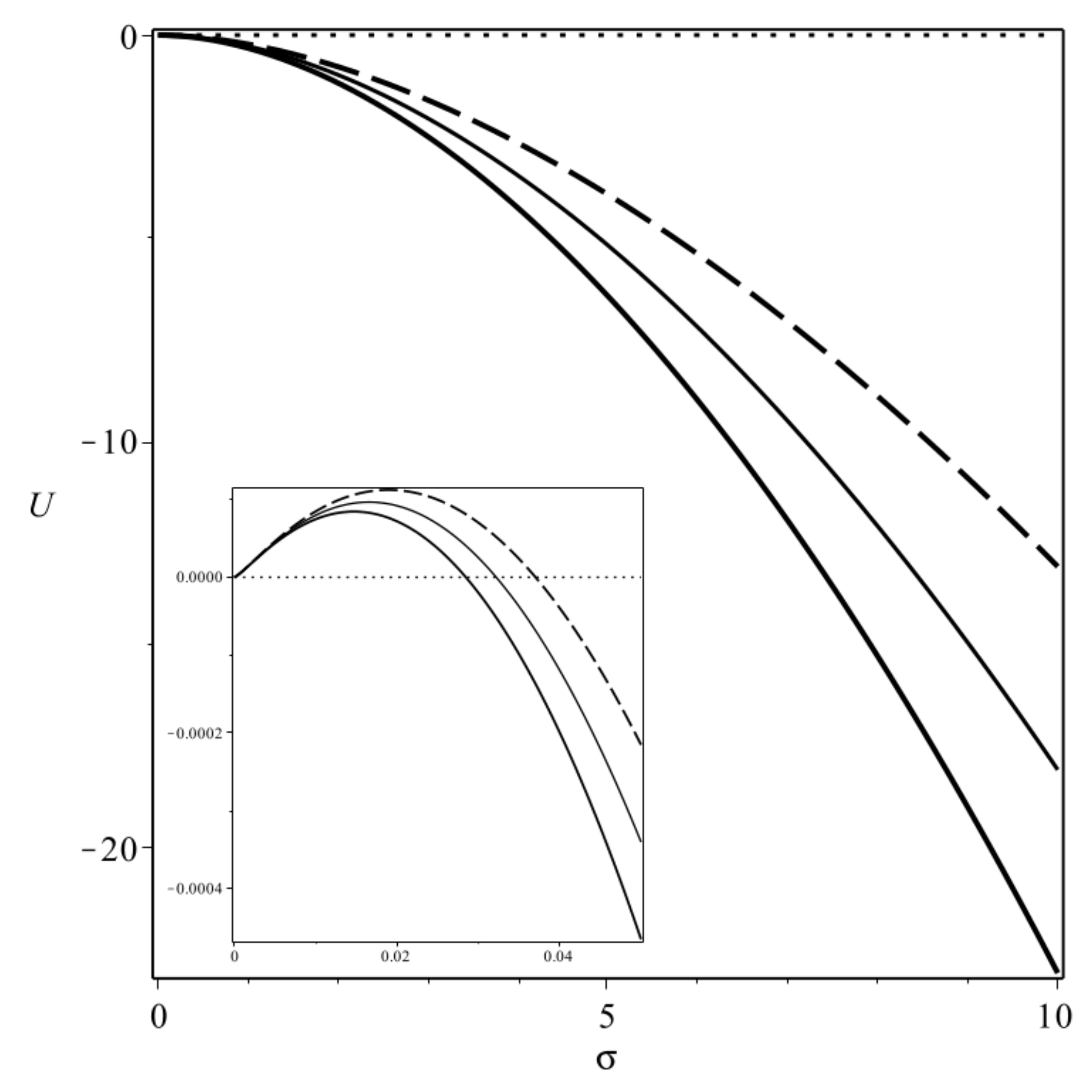}
\caption{The effective potential \eqref{veffc1} depicted for $a=4/9$, $s=3$, and $\omega^2={0.6},{0.65}$ and $0.7$ (top) and $\omega^2=1,{1.1}$ and $1.2$ (bottom), with the dashed line standing for the case $\omega=1$. In the two figures, the thickness of the lines increases with $\omega$, and the insets show the behavior near $\sigma=0$.}
\label{fig2}
\end{figure}

To investigate compact Q-balls we get motivation from the work \cite{cl} and consider the following family of models in $(1,1)$ dimensions
\be\label{new}
{\cal L} = \frac12 \partial_\mu\vphi^* \partial^\mu \vphi - \frac12 |\vphi|^2 + \frac13|\vphi|^{2-1/s} - \frac14 a\,|\vphi|^{2-2/s},
\ee
where $s$ is a real parameter restricted to obey $s>2$. In this case, we use the standard route and write effective potential \eqref{veff} in the form
\be\label{veffc1}
U(\sigma) = \frac12(1-\omega^2) \sigma^2 - \frac13\sigma^{2-1/s} + \frac14 a\,\sigma^{2-2/s}\,.
\ee
From the above expression, we see that the effective potential only has a global minimum if $\omega^2<1$, and we depict it for some values of $a, s$ and $\omega$ in Fig.~\ref{fig2}.

As we can see, the non vanishing zero of $U(\sigma)$ is given by
\be
\bar{\sigma} = \left(2\frac{1+\sqrt{1-9(1-\omega^2)a/2}}{3a}\right)^{-s}.
\ee
Also, the equation of motion for $\sigma(x)$ has the form
\ben\label{eqm}
\!\!\!\!\sigma^{\prime\prime} \!=\! (1\!-\!\omega^2)\sigma \!-\! \frac13\! \left(2\!-\!\frac{1}{s}\right)\sigma^{1-\frac{1}{s}}\! +\!\frac{a}{2}\! \left(1\!-\!\frac{1}{s}\right)\sigma^{1-\frac{2}{s}}\,.
\een
This is the equation we should solve, and it now depends upon $a$, $s$, and $\omega$. The condition \eqref{condomega1} is still valid, and gives $\omega_-=\sqrt{1-2/(9a)}$ and $\omega_+=\infty$ in this case. Again, we take $a\geq 2/9$ to assure that $\omega$ is real. Then, the solutions of \eqref{eqm} are valid for any $\omega >\omega_{-} $.

\subsection{Analytical solutions}

We consider the case $\omega_-<\omega<1$. The equation \eqref{eqm} is solved by
\be\label{solcomp1}
\sigma(x) =\begin{cases}
 {\left(\frac{1-\omega^2}{2a}\right)}^{-s/2}\left[\coth\left(\frac{\sqrt{1-\omega^2}}{2s}\, x + b \right) \right. & \\
 \left.-\coth\left(\frac{\sqrt{1-\omega^2}}{2s}\, x - b \right)\right]^{-s}, & |x|\leq x_0 \\
 0, & |x|>x_0,
\end{cases}
\ee
where $x_0=2sb/\sqrt{1-\omega^2}$ is the width of the solution at half height, and $b$ is a constant, given by
\be
b=\text{arccoth}\left({\sqrt{2}}\; \frac{1+\sqrt{1-9(1-\omega^2)a/2}}{3\sqrt{a(1-\omega^2)}} \right)\,.
\ee

The next case is for $\omega=1$. Here we get the solution
\be\label{solcomp2}
\sigma(x) =\begin{cases}
 \left( \frac{3a}{4} -\frac{x^2}{6s^2}\right)^s, & |x|\leq x_0 \\
 0, & |x|>x_0,
\end{cases}
\ee
where $x_0$ now changes to $x_0=3s\sqrt{a/2}$.

\begin{figure}[t!]
\includegraphics[width=6.4cm,height=5.5cm]{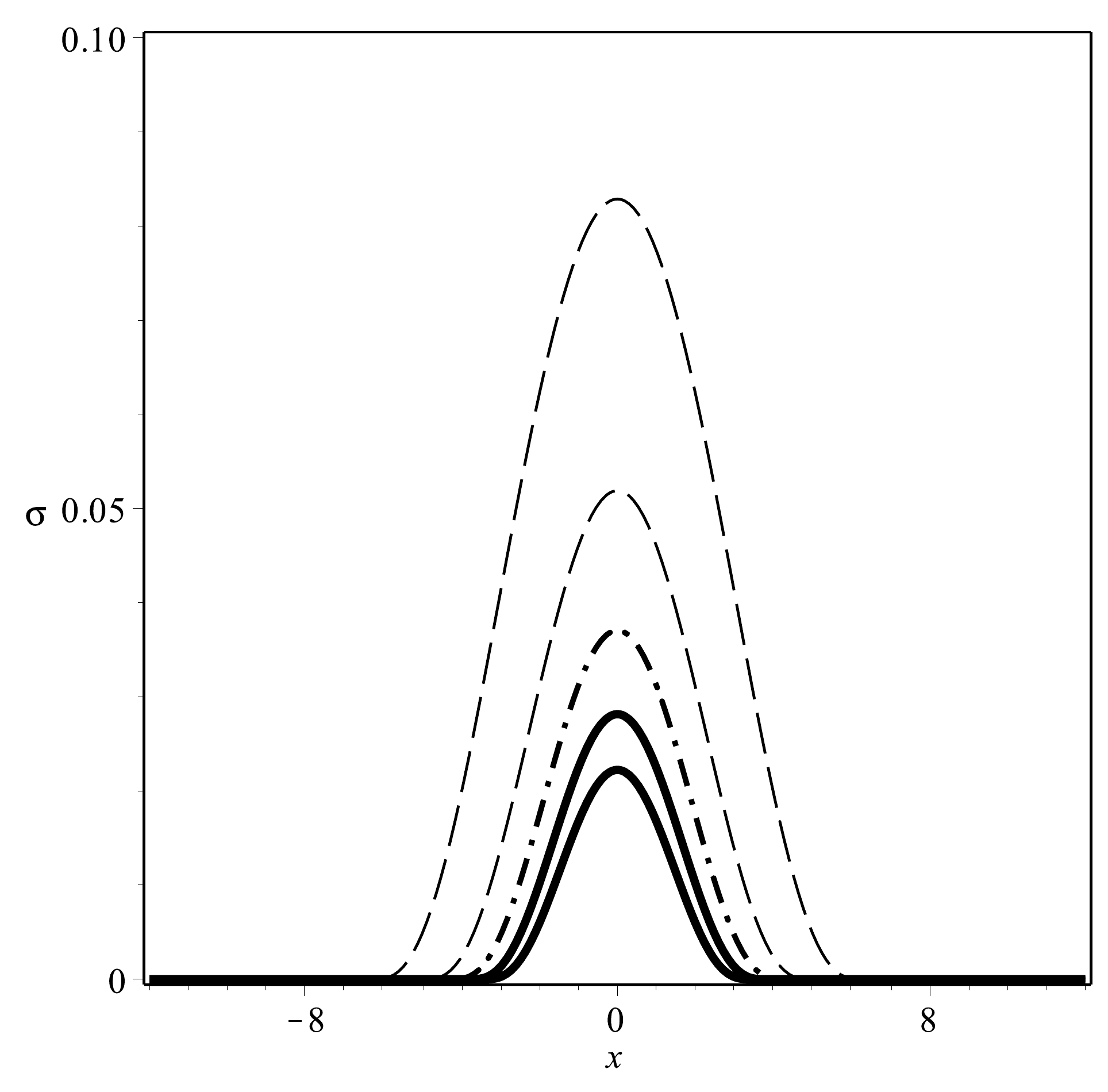}
\caption{The solutions \eqref{solcomp1}, \eqref{solcomp2} and \eqref{solcomp3} depicted for $a=4/9$ and $s=3$, with $\omega=0.8,0.9,1,1.1$ and $1.2$. The thickness of the lines increases with $\omega$. The dashed lines represent the solution \eqref{solcomp1}, for $\omega<1$, the dash-dotted line stands for the solution \eqref{solcomp2}, for $\omega=1$, and the solid lines are for the solution \eqref{solcomp3}, for $\omega>1$.}
\label{fig3}
\end{figure}

\begin{figure}[t!]
\includegraphics[width=6.4cm,height=5.5cm]{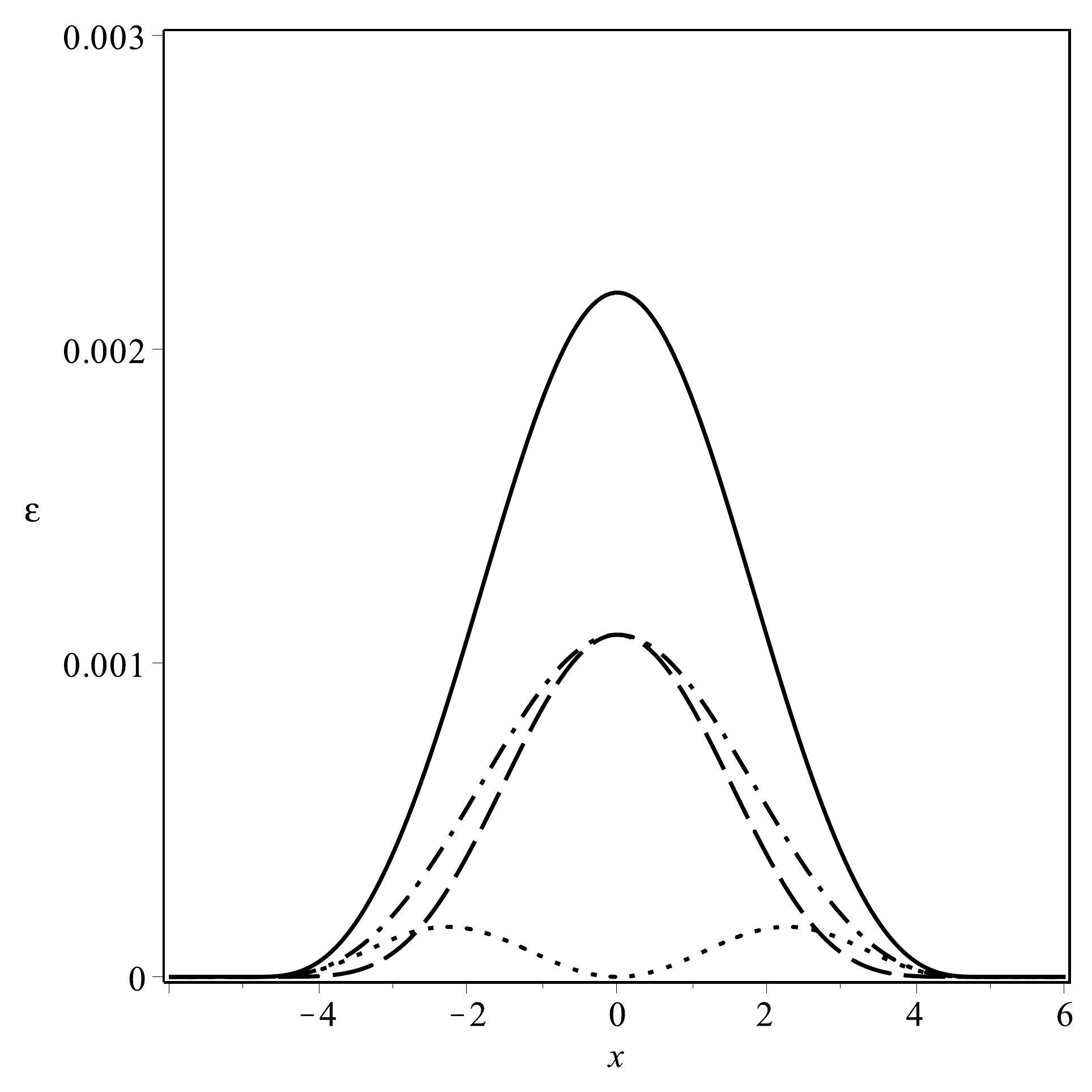}
\caption{The total energy density, depicted with a solid line for $a=4/9$, $s=3$, and $\omega=0.9$. We also show the kinetic, gradient and potential energy densities, depicted with dashed, dot-dashed and dotted lines, respectively.}
\label{fig4}
\end{figure}
\begin{figure}[t!]
\includegraphics[width=6.0cm,height=5.5cm]{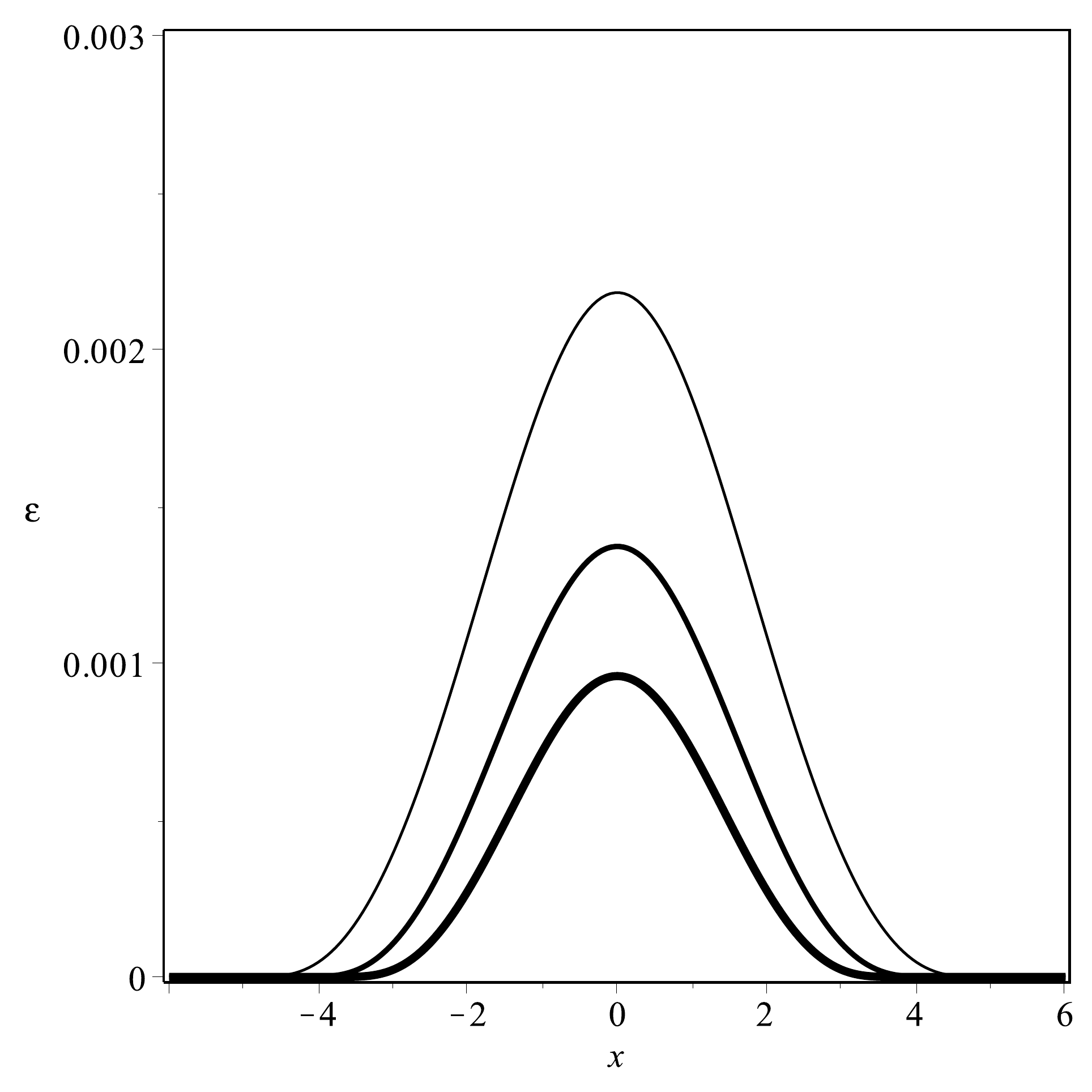}
\caption{The total energy density depicted for $a=4/9$, $s=3$, and $\omega={0.9},{1}$ and $1.1$, with the thickness of the lines increasing with increasing $\omega$. We note that both the amplitude and width of the total energy density decrease, as we increase $\omega$.}
\label{fig5}
\end{figure}

The last case is for $\omega>1$, and now the equation \eqref{eqm} is solved by
\be\label{solcomp3}
\sigma(x) =\begin{cases}
 {\left(\frac{\omega^2-1}{2a}\right)}^{-s/2}\left[\cot\left(\frac{\sqrt{\omega^2-1}}{2s}\, x + d \right) \right. & \\
 \left.-\cot\left(\frac{\sqrt{\omega^2-1}}{2s}\, x - d \right)\right]^{-s}, & |x|\leq x_0 \\
 0, & |x|>x_0,
\end{cases}
\ee
where $x_0=2sd/\sqrt{\omega^2-1}$ and $d$ is given by
\be
d=\text{arccot}\left(\sqrt{2}\;\frac{1+\sqrt{1+9(\omega^2-1)a/2}}{3\sqrt{a(\omega^2-1)}}\right).
\ee
We depict these solutions \eqref{solcomp1}, \eqref{solcomp2} and \eqref{solcomp3} in Fig.~\ref{fig3} for some values of
$\omega$. The figure shows that the amplitude becomes smaller as we increase $\omega$ and more importantly, that they all vanish, outside a compact interval of the real line. One should compare this behavior with the behavior that appears in Fig.~\ref{fig1} for the standard Q-balls, which shows that the solutions vanish asymptotically. Thus, differently from the standard Q-balls, these new compact Q-balls do not have a tail, so they seem to behave as hard charged balls.

\subsection{Charge and energy}

Due to the importance of the above new results on Q-balls, we further investigate their main characteristics. In particular, we have calculated the charge for a general $s$. It is given by
\ben\label{generalcharge}
&&Q=\frac{2\,s\,\omega\sqrt{2\pi}\,3^{2s+1}a^{2s+1/2}}{\left( 2 \left(1+\sqrt{1-9(1-\omega^2)a/2}\right)\right)^{2s+1}}\frac{\Gamma(2s+1)}{\Gamma\left(2s+\frac32\right)} \nonumber \\
&& _2F_1\left(\frac12,2s+1;2s+\frac32;\frac{9(1-\omega^2)a/2}{\left(1+\sqrt{1-9(1-\omega^2)a/2}\right)^2}\right), \nonumber\\
\een
where $\Gamma(z)$ is Gamma Function and $_2F_1\left(a,b;c;z\right)$ is the Hypergeometric Function. Specifically, for $a=2/9$ we have $\omega_-=0$, and this makes $Q\to0$ if $\omega\to\omega_-=0$. For $a>2/9$ we have $Q\to\infty$ if $\omega\to\omega_-$. Moreover, the expression for the energy density is cumbersome, so we omit it here. However, we depict in Figs.~\ref{fig4} and
\ref{fig5} the energy densities for several values of $a$, $s$ and $\omega$. 

As shown before, the kinetic energy is given in terms of the charge \eqref{generalcharge} as in Eq.~\eqref{ke}. The gradient energy is given by
\ben
\!\!\!\!\!\!\!\!\!\!\!\!&&\!\!\!\!\!\!\!\!E_g\! =\!\frac{\sqrt{\pi/2}\;3^{2s-1}a^{2s-1/2} s(s-1)}{\left(2\left(1\!+\!\sqrt{1\!-\!9(1\!-\!\omega^2)a/2}\right)\right)^{2s-1}} \frac{\Gamma(2s\!-\!2)}{\Gamma\left(2s\!+\!\frac12\right)}\;\;\;\nonumber\\
\!\!\!\!\!\!\!\!&&\!\!\!\!\!\!\!\! _2F_1\!\!\left(\!\!-\frac12,2s\!-\!1;2s+\frac12;\frac{9(1\!-\!\omega^2)a/2}{\left(1\!+\!\sqrt{1\!-\!9(1\!-\!\omega^2)a/2}\right)^2}\!\!\right)\!\!.\;\;\; \!\!\!
\een

\subsection{Stability}

\begin{figure}[t!]
\includegraphics[width=5.6cm,height=4.2cm]{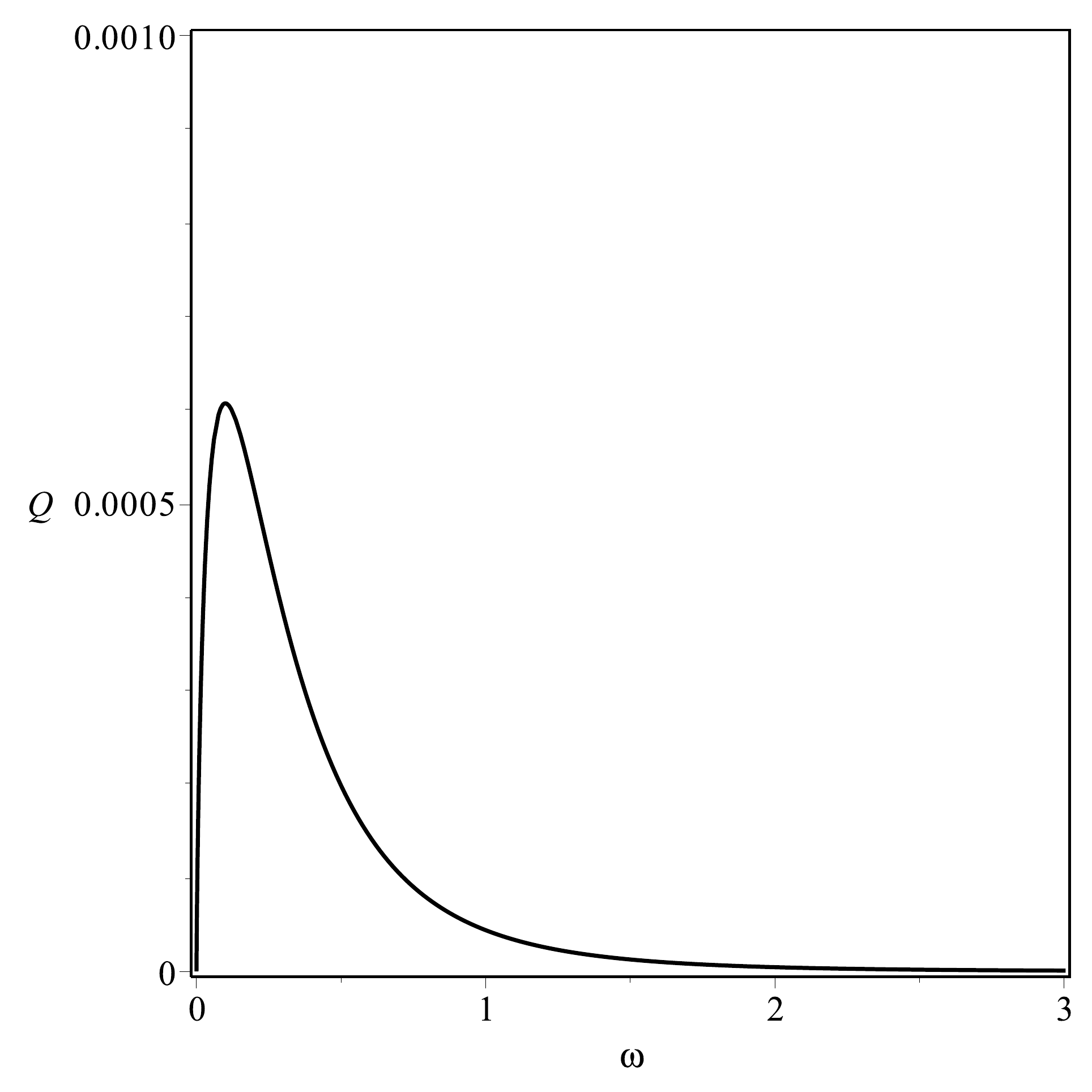}
\includegraphics[width=5.6cm,height=4.2cm]{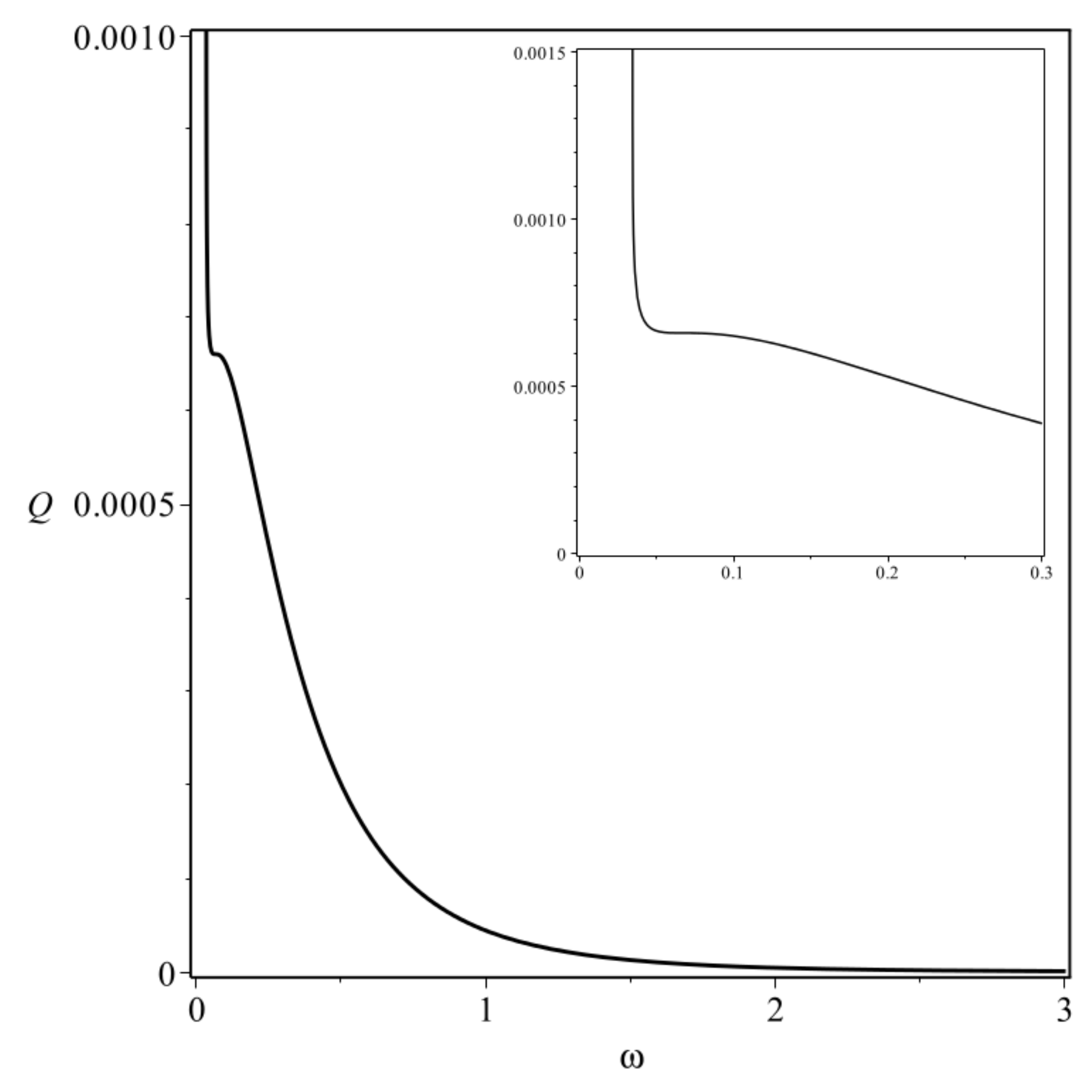}
\includegraphics[width=5.6cm,height=4.2cm]{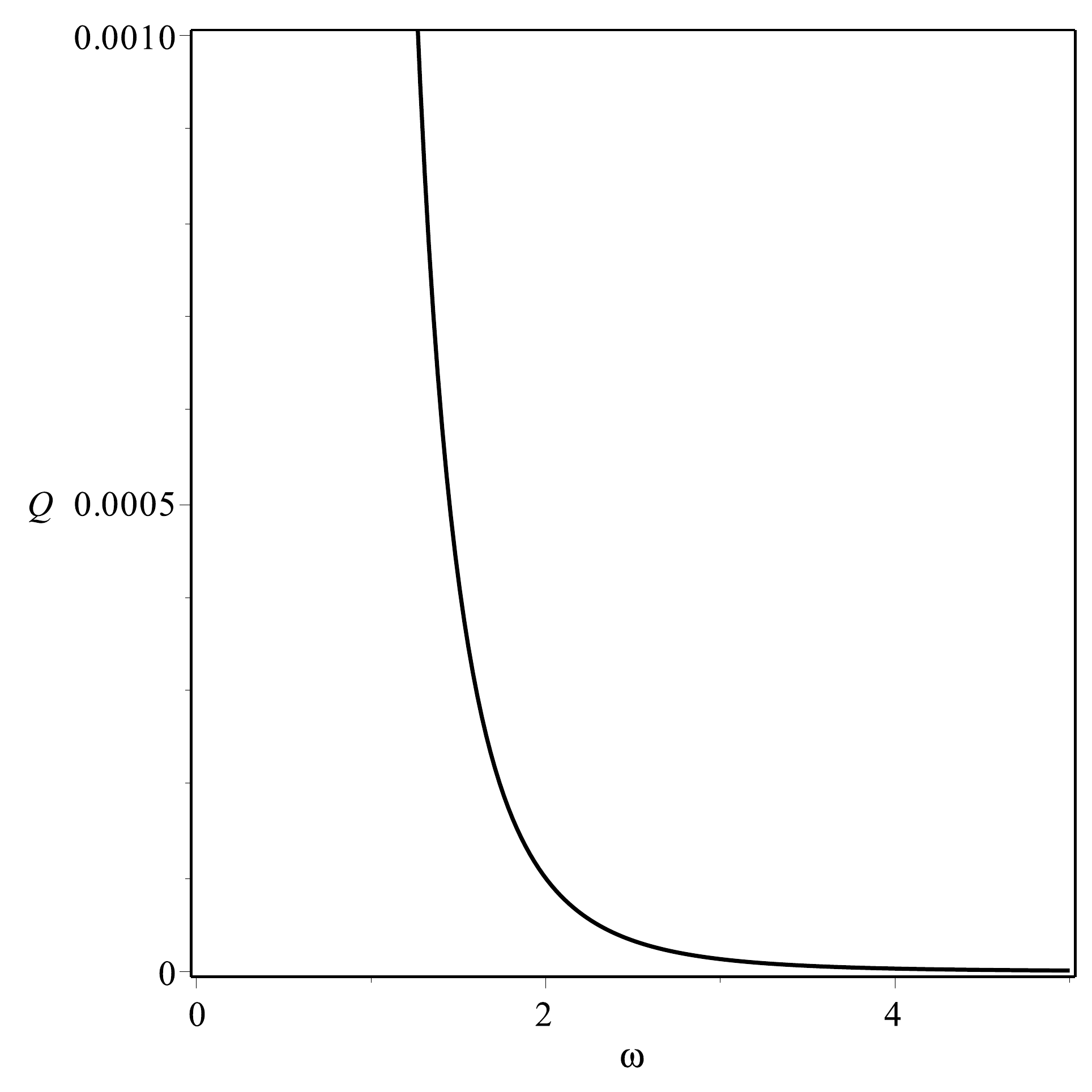}
\caption{The charge \eqref{generalcharge} for $s=3$ and $a=a_0=2/9$ (top), $a=a_1=0.22248865$ (middle) and $a=4/9$ (bottom). The inset in the middle panel amplifies the behavior of the charge for $\omega$ near $\omega_-$. }
\label{fig6}
\end{figure}

We now focus on the stability of the Q-balls. We follow the lines of \cite{bmm} to see that the model is quantum mechanically stable, because $\omega_+$ is infinity and then the relation $E/Q<\omega_+=\infty$ is satisfied for $a\geq2/9$. To study classical stability, we have to show that the charge $Q$ is a monotonically decreasing function of $\omega.$ To see this, we start with $a=a_0\equiv2/9=0.22222222$. At this value, the charge is not monotonically decreasing. We then increase $a$ up to $a=a_1\equiv 0.22248865$, the value where the charge starts to become monotonically decreasing with $\omega$, so that the Q-balls become classically stable for $a>a_1$. We have noted that for $2/9<a<a_1$ the charge $Q$ is not a single valued function of $\omega$, but for $a>a_1$ it decreases monotonically. We illustrate this in Fig.~\ref{fig6}, where we depict the charge as a function of $\omega$ for $a=a_1$, with the two insets showing how it varies for $a=2/9$, and for $a=4/9$. 

We then conclude that the compact Q-balls solutions that we found above are stable quantum mechanically for $a\geq2/9=0.22222222$, and that they are classically stable for $a$ greater than $a_1=0.22248865$.

\section{Ending comments}
\label{sec:end}

In this work we investigated the construction of compact Q-balls in models described by a single scalar field in $(1,1)$ spacetime dimensions. We started briefly reviewing the case of standard Q-balls, following the lines of \cite{bmm}. Inspired by results obtained before in \cite{cl}, where one describes the presence of compact lumps, we then constructed a model, defined in Eq.~\eqref{new}, which can be used to describe Q-ball solutions of the compact type, with the solution being non trivial only in a compact interval of the real line.

We used the analytical solutions to describe the charge and energy of the compact Q-balls, which we further explored to study stability. The results show how to control the parameters $a$ and $s$ to construct compact Q-balls that are stable, classically and quantum mechanically. In this context, the compact Q-balls seem to appear as hard charged balls, so they may have different collective behavior. Also, they may act differently, if they live in a space in the presence of compact extra dimensions. 

We believe that the results of the work are of current interest, so we shall further explore similar issues in another work, paying closer attention to the case of the gauged version of the model here investigated, and also, in higher spatial dimensions. There are several possible directions of investigations, some of them may follow as in the recent works \cite{qbrane,gqb,jap}. In particular, the role played by gauged compact Q-balls, although beyond the present scope, shall be estimated upon different baryon and lepton charges \cite{jap}. In addition,  as solutions of the supersymmetric extension Standard Model, Q-balls were studied, and their compact gravitating counterparts should also be studied \cite{Brihaye:2014gua}.

{\bf{Acknowledgements.}} The authors are grateful to Brazilian agencies for partial financial support. DB thanks support from CNPq grants 455931/2014-3 and 06614/2014-6, LL thanks support from CNPq grants 307111/2013-0 and 447643/2014-2, MAM thanks support from CNPq grant 140735/2015-1, RM thanks support from CNPq grants 508177/2010-3 and 455619/2014-0, and RdR thanks support from CNPq grants 473326/2013-2 and 303293/2015-2, and also from FAPESP grant 2015/10270-0.

\end{document}